\documentstyle[aps,epsf]{revtex}
\begin{document}
\draft
\tighten


\title{Limits on R--parity non--conservation in 
MSSM with gauge mediated supersymmetry breaking 
from the neutrinoless double beta decay}
\author{ 
A. Wodecki, Wies{\l}aw A. Kami{\'n}ski
}
\address{
Theoretical Physics Department, Maria Curie--Sk{\l}odowska University, \\
Radziszewskiego 10, 20--031 Lublin, Poland}
\date{\today}
\maketitle


\begin{abstract}
The R-parity non--conservation phenomenology 
within the minimal supersymmetric standard model 
with gauge mediated supersymmetry breaking
has been studied. New constraints on the lepton number
violating constant 
$\lambda _{111}^\prime$ were imposed by 
non--observability of the neutrinoless double beta decay.
We find that $\lambda _{111}^\prime$ depends
strongly on the effective supersymmetry breaking scale
$\Lambda$ only and deduce limits which are much stronger than those
previously found in literature.
\end{abstract}
\pacs{12.60.Jv, 11.30.Er, 23.40.Bw}


\newcommand{\newc}{\newcommand}
\newc{\be}{\begin{equation}}
\newc{\ee}{\end{equation}}
\newc{\bea}{\begin{eqnarray}}
\newc{\eea}{\end{eqnarray}}
\newc{\Lsoft}{{\cal L}_{soft}}
\newc{\hinot}{\widetilde H^0_2}
\newc{\hinob}{\widetilde H^0_1}
\newc{\bino}{\widetilde B^0}
\newc{\wino}{{\widetilde W^0_3}}
\newc{\gluino}{{\widetilde g}}
\newc{\mgluino}{m_{\gluino}}
\newc{\msl}{m_{\widetilde l}}
\newc{\msq}{m_{\widetilde q}}
\newc{\msf}{m_{\widetilde f}}
\newc{\mz}{m_Z}
\newc{\mw}{m_W}
\newc{\superhu}{\hat H_u}	\newc{\superhd}{\hat H_d}
\newc{\superl}{\hat L}		\newc{\superr}{\hat R}
\newc{\superec}{\hat {e}^c}
\newc{\superq}{\hat Q}
\newc{\superu}{\hat U}
\newc{\superd}{\hat D}
\newc{\superuc}{\hat {u}^c}
\newc{\superdc}{\hat {d}^c}
\newc{\tildeQ}{\widetilde Q}
\newc{\tildeU}{\widetilde U}
\newc{\tildeD}{\widetilde D}
\newc{\tildeL}{\widetilde L}
\newc{\tildeuc}{\widetilde {u}^c}
\newc{\tildedc}{\widetilde {d}^c}
\newc{\tildeec}{\widetilde {e}^c}
\newc{\tildenu}{\widetilde \nu}
\newc{\MS}{{\rm\overline{MS}}}
\newc{\DR}{{\rm\overline{DR}}}
\newc{\stopq}{{\widetilde t}}		\newc{\stp}{\stopq}
\newc{\stopl}{\widetilde t_L}
\newc{\stopr}{\widetilde t_R}
\newc{\stopone}{\widetilde t_1}
\newc{\stoptwo}{\widetilde t_2}

\newc{\mstopq}{m_{\tilde t}}
\newc{\mstopl}{m_{\tilde t_L}}
\newc{\mstopr}{m_{\tilde t_R}}
\newc{\mstopone}{m_{\tilde t_1}}
\newc{\mstoptwo}{m_{\tilde t_2}}
\newc{\stau}{{\widetilde \tau}}
\newc{\staul}{\widetilde \tau_L}
\newc{\staur}{\widetilde \tau_R}
\newc{\stauone}{\widetilde \tau_1}
\newc{\stautwo}{\widetilde \tau_2}

\newc{\mstau}{m_{\tilde \tau}}
\newc{\mstaul}{m_{\tilde \tau_L}}
\newc{\mstaur}{m_{\tilde \tau_R}}
\newc{\mstauone}{m_{\tilde \tau_1}}
\newc{\mstautwo}{m_{\tilde \tau_2}}
\newc{\onehalf}{\frac{1}{2}}
\newc{\vhiggs}{V_{\rm Higgs}}
\newc{\delV}{\Delta V}
\newc{\sinsqthw}{\sin^2\theta_{\rm w}}
\newc{\ai}{\alpha_1}
\newc{\aii}{\alpha_2}
\newc{\ev}{{\rm\,eV}}
\newc{\mev}{{\rm\,MeV}}
\newc{\gev}{{\rm\,GeV}}
\newc{\tev}{{\rm\,TeV}}
\newc{\agut}{\alpha_X}
\newc{\Uy}{\mbox{U(1)}_{\rm Y}}
\newc{\Uem}{\mbox{U(1)}_{\rm EM}}
\newc{\vev}{\hbox{\it v.e.v.}}
\newc{\vone}{v_d}	\newc{\vtwo}{v_u}
\newc{\sbot}{{\tilde{b}}}
\newc{\slepton}{{\tilde{l}}}
\newc{\supq}{{\tilde{u}}}
\newc{\sdown}{{\tilde{d}}}
\newc{\selectron}{{\tilde{e}}}
\newc{\sneutrino}{{\tilde{\nu}}}
\newc{\squark}{{\tilde{q}}}
\newc{\higgsino}{{\tilde{H}}}
\newc{\bsg}{b\to s\gamma}
\newc{\mh}{\mhl}
\newc{\mx}{\mgut}
\newc{\eL}{\tilde{e}_L}
\newc{\eR}{\tilde{e}_R}
\newc{\Rslash}{\not R}

Gauge--mediated theories of supersymetry (SUSY) breaking (GMSB) 
have attracted a great  deal of attention recently because of their 
high predictability, a natural solution of the flavour problem, 
a new phenomenology and much less free parameters compared to 
the Minimal Supersymmetric Standard Model (MSSM), where SUSY 
breaking is mediated by the gravitational 
interaction \cite{gmth1,gmth2,gmth3,martin,gmth4,gmphen1,gmphen2,haberkane}.
In the GMSB models SUSY breaking is transmitted to the superpartners 
of quarks, leptons and gauge bosons via the usual 
$SU(3) \times SU(2) \times U(1)$ gauge interactions and occurs at 
the scale of the order $M_{SUSY}\sim 10^5$ GeV. 
Gauginos and sfermions acquire their masses through interactions 
with the messenger sector at the one-- and two--loop levels
respectively, resulting in different phenomenology from the MSSM one
of the low--energy world. 
In these models flavour--diagonal sfermions 
mass matrices are induced in a rather low energy scale, therefore 
they supply us with a very natural mechanism of suppressing flavour      
changing neutral currents (FCNC). 
Moreover, since the soft masses arise as gauge charges squared, the 
sizeable hierarchy proportional to the gauge quantum numbers appears 
among the superpartner masses.
As in MSSM the electroweak symmetry breaking (EWSB) is also driven 
by negative radiative corrections to the up--type Higgs mass resulting        
from the large top--quark Yukawa coupling and stop masses. 
Thus, with EWSB constraints the minimal GMSB models are highly 
constrained and strongly predictive. 
Recently renewed interest in GMSB \cite{gmphen1,gmphen2} is then
understood.

Motivated by the above features of the GMSB models we study the 
R--parity breaking phenomenology of MSSM and propose to use non--standard
processes like the neutrinoless double beta decay ($0\nu \beta \beta $)
for deducing the limits on some R-parity breaking constants. 
In the previous studies such estimates were performed in the framework 
of MSSM with supergravity mediated SUSY breaking
by means of additional assumptions relating sfermion and gauginos masses
\cite{koval1,koval2} or using the GUT constraints \cite{wodecki}.
Within the GMSB models one can find quantitatively new constraints, which 
may serve as a hint for further accelerator searches of R--parity breaking.   

The R--parity imposed on MSSM can be explicitely violated by the 
trilinear \cite{rbreaking} and bilinear  \cite{valle} terms in the 
superpotential. 
The trilinear terms lead to the lepton number and flavour violation, 
while the bilinear terms generate the non--zero vacuum 
expectation values for the sneutrino fields $\langle\tilde\nu_l
\rangle$, causing neutrino--neutralino mixing and 
electron--chargino mixing. 
Due to the lepton number violation, supersymmetric models with 
the R--parity broken may serve as a basis for description of some
exotic processes. 
In the case of non--observability of $0\nu \beta \beta $ one can 
impose limits on the R--parity breaking  parameters. 
It has been found that $0\nu \beta \beta $ is a powerful tool 
for such constraints \cite{koval1,koval2,wodecki}. 

To obtain the effective Lagrangian of the  $0\nu \beta \beta $
decay we have to write the complete superpotential including both 
R--parity conserving and breaking parts:
\be
W = W_0 + W_{\Rslash}
\label{spotential:eq}
\ee
with
\be
W_0 = h^U_{ij} \superq_i\superhu\superuc_j
           + h^D_{ij} \superq_i\superhd\superdc_j
	   + h^E_{ij} \superl_i\superhd\superec_j
	   + \mu \superhd\superhu.
\label{spotential_0:eq}
\ee
and
\be
W_{\Rslash} = \lambda_{ijk} \superl_i\superl_j\superuc_k
           + \lambda_{ijk}'  \superl_i\superq_j\superdc_k
	   + \lambda_{ijk}''  \superuc_i\superdc_j\superdc_k.
\label{spotential_R:eq}
\ee
In Eqs. (\ref{spotential_0:eq})--(\ref{spotential_R:eq}) 
$ \superq, \superl$ denote the quark and lepton SU(2) 
doublet superfields and $ \superuc, \superdc, \superec $ are 
corresponding SU(2) singlets. 
The Higgs superfields $ \superhu, \superhd $ contain scalar 
components giving mass to the up-- and down--type quarks and leptons. 
In the R--parity breaking part we set $\lambda_{ijk} = \lambda_{ijk}''
= 0$ to avoid the unsuppressed proton decay.  
Since supersymmetry in the low--energy world is broken, the Lagrangian of
the theory is supplemented with the "soft" supersymmetry breaking terms:
\bea
-\Lsoft
& = & \left(A^U_{(ij)} h^U_{ij} \tildeQ_i H_u \tildeuc_j
    + A^D_{(ij)} h^D_{ij} \tildeQ_i H_d \tildedc_j
    + A^E_{(ij)} h^E_{ij} \tildeL_i H_d \tildeec_j + h.c. \right)  
    \nonumber \\
&   & \mbox{} + B\mu \left(H_d H_u + h.c. \right)
    + m_{H_d}^2\vert H_d\vert^2 + m_{H_u}^2\vert H_u\vert^2 \nonumber\\
&   & \mbox{} + m_{\tildeL}^2\vert \tildeL\vert^2 + m_{\tildeec}^2\vert
\tildeec\vert^2
     + m_{\tildeQ}^2\vert \tildeQ\vert^2 + m_{\tildeuc}^2\vert \tildeuc\vert^2
     + m_{\tildedc}^2\vert \tildedc\vert^2 \nonumber \\
&   & \mbox{} + \left(\onehalf M_1 \bar{\psi}_B\psi_B + \onehalf
	M_2\bar{\psi}^a_W
	\psi^a_W + \onehalf \mgluino\bar{\psi}^a_g\psi^a_g + h.c. \right),
\label{lsoft:eq}
\eea
where a tilde denotes the scalar partners of quark and lepton fields, 
while  $\psi_i$'s are the spin--$1 \over 2$ partners of the gauge bosons. 

Using the R--breaking term $W_{\Rslash}$ one can find the lepton 
number violating Lagrangian.
\begin{eqnarray}
  {\cal L}_{\lambda'_{111}}=
  -\lambda'_{111}[({\bar{u}_{L}},{\bar{d}_{L}})
     \left(
         \begin {array}{c}
           e_{R}^{c}\\-\nu_{R}^{c}
         \end{array}
     \right)
  \tilde{d}_{R}^{\ast}+({\bar{e_{L}}},{\bar{\nu_{L}}}) d_{R}
     \left(
         \begin {array}{c}
           \tilde{u}^{\ast}_{L} \\ -\tilde{d}^{\ast}_{L}
         \end{array}
     \right)
 + \\
  ({\bar{u_{L}}},{\bar{d_{L}}}) d_{R}
     \left(
         \begin {array}{c}
           \tilde{e}^{\ast}_{L} \\ -\tilde{\nu}^{\ast}_{L}
         \end{array}
     \right)
 + h.c.]\,. \nonumber
  \end{eqnarray}
Together with  Lagrangians describing interactions among
gluinos, neutralinos, fermions and sfermions \cite{haberkane} 
we end up, after integration out of heavy degrees of freedom,  
with the final effective Lagrangian
\begin{eqnarray}
 {\cal L}^{\Delta L_e =2}_{eff}\ =
\frac{G_F^2}{2 m_{_p}}~ \bar e (1 + \gamma_5) e^{\bf c} 
\left[\eta_{PS}~J_{PS}J_{PS} 
- \frac{1}{4} \eta_T\   J_T^{\mu\nu} J_{T \mu\nu} \right],
\label{susy.2}
\end{eqnarray}
where the color singlet hadronic currents\footnote{At this point we
want to stress a need for a proper treatment of the colour currents in 
Lagrangian (\ref{susy.2})\cite{wodecki}.
In the previous papers \cite {koval1} this Lagrangian is erroneous.}
are $J_{PS} = {\bar u}^{\alpha} \gamma_5 d_{\alpha} + 
{\bar u}^{\alpha} d_{\alpha}$, 
$J_T^{\mu \nu} = {\bar u}^{\alpha} 
\sigma^{\mu \nu} (1 + \gamma_5) d_{\alpha}$, 
with  $\alpha$ as a color index and 
$\sigma^{\mu \nu} = (i/2)[\gamma^\mu , \gamma^\nu ]$.
The effective lepton--number violating parameters $\eta_{PS}$
and $\eta_{T}$  in Eq.\ (\ref{susy.2}) accumulate fundamental parameters
of  MSSM:
\begin{eqnarray}
\label{etaq}
\eta_{PS} &=&  \eta_{\chi\tilde e} + \eta_{\chi\tilde f} +
\eta_{\chi} + \eta_{\tilde g} + 7 \eta_{\tilde g}^{\prime}, \\
\label{eta}
\eta_{T} &=& \eta_{\chi} - \eta_{\chi\tilde f} + \eta_{\tilde g}
- \eta_{\tilde g}^{\prime},
\end{eqnarray}
where
\begin{eqnarray*}
\eta_{\tilde g} = \frac{\pi \alpha_s}{6}
\frac{\lambda^{'2}_{111}}{G_F^2 m_{\tilde d_R}^4} \frac{m_P}{m_{\tilde 
g}}\left[
1 + \left(\frac{m_{\tilde d_R}}{m_{\tilde u_L}}\right)^4\right], &
\,\,\,\,\,\,\,\,\,\,
\eta_{\chi} = \frac{ \pi \alpha_2}{2}
\frac{\lambda^{'2}_{111}}{G_F^2 m_{\tilde d_R}^4}
\sum_{i=1}^{4}\frac{m_P}{m_{\chi_i}}
\left[
\epsilon_{R i}^2(d) + \epsilon_{L i}^2(u)
\left(\frac{m_{\tilde d_R}}{m_{\tilde u_L}}\right)^4\right],
\end{eqnarray*}

\begin{eqnarray*}
\eta_{\chi \tilde e} = 2 \pi \alpha_2
\frac{\lambda^{'2}_{111}}{G_F^2 m_{\tilde d_R}^4}
\left(\frac{m_{\tilde d_R}}{m_{\tilde e_L}}\right)^4
\sum_{i=1}^{4}\epsilon_{L i}^2(e)\frac{m_P}{m_{\chi_i}},&
\,\,\,\,\,\,\,\,\,\,
\eta'_{\tilde g} = \frac{\pi \alpha_s}{12}
\frac{\lambda^{'2}_{111}}{G_F^2 m_{\tilde d_R}^4}
\frac{m_P}{m_{\tilde g}}
\left(\frac{m_{\tilde d_R}}{m_{\tilde u_L}}\right)^2,
\end{eqnarray*}

\begin{eqnarray}
\label{eta_end}
\eta_{\chi \tilde f} = \frac{\pi \alpha_2 }{2}
\frac{\lambda^{'2}_{111}}{G_F^2 m_{\tilde d_R}^4}
\left(\frac{m_{\tilde d_R}}{m_{\tilde e_L}}\right)^2
\sum_{i=1}^{4}\frac{m_P}{m_{\chi_i}}
\left[\epsilon_{R i}(d) \epsilon_{L i}(e) \right.
+ \left.\epsilon_{L i}(u) \epsilon_{R i}(d)
\left(\frac{m_{\tilde e_L}}{m_{\tilde u_L}}\right)^2
+ \epsilon_{L i}(u) \epsilon_{L i}(e)
\left(\frac{m_{\tilde d_R}}{m_{\tilde u_L}}\right)^2
\right]. 
\end{eqnarray}

So, the effective Lagrangian depends on many supersymmetric parameters.
In order to find their low--energy values in the GMSB model, 
we proceed as follows.

The minimal model of GMSB considered in this paper consists of messenger
fields which transform as single flavor of $5+\bar{5}$ of $SU(5)$, i.e.
they are $SU(2)_L$ doublets $l$ and $\tilde{l}$ and $SU(3)$ triplets $q$
and $\tilde{q}$. 
The $SU(3)\times SU(2)\times U(1)$ gauge interactions of
those fields communicate supersymmetry breaking from a hidden sector to the
fields of the visible world via coupling to a gauge singlet S in the
superpotential
\begin{equation}
W=\lambda _2Sl\tilde{l}+\lambda _3Sq\tilde{q}.
\end{equation}
The lowest and F--components of the singlet superfield S acquire vev's and
set the overall scale for the messenger sector and SUSY breaking
respectively. 
For $F\neq 0$ the messenger sector loses its supersymmetry: 
\begin{eqnarray}
m_b&=&M\sqrt{1\pm \frac \Lambda M}\,, \\
m_f&=&M\,,
\end{eqnarray}
where $M=\lambda S$, $\Lambda =\frac FS$.

The messenger fields transmit the SUSY breaking to the visible sector
through loops containing insertions of S and results with the 
gaugino and scalar masses:
\be
 m_{\lambda _i}(M)=N_m g(\frac{\Lambda}{M})
 \frac{\lambda _i(M)}{4\pi }\Lambda \,,  
\label{eq1}  
\ee
\be
m^2(M)=2N_mf(\frac{\Lambda}{M})\sum_{i=1}^3k_i\left
( \frac{{\lambda _i(M)}}{{4\pi }}
\right) ^2  \lambda ^2\,.
\label{eq2}
\ee
In Eq. (\ref{eq2}) we sum up over $SU(3)\times SU(2)_L\times U(1)_Y$ 
with $k_1=%
\frac 35(\frac Y2)^2,Y=2(Q-T_3),k_2=\frac 34$ for $SU(2)_L$ doublets and
zero for singlets, and $k_3=\frac 43$ for $SU(3)_C$ triplets and zero
for singlets. 
$N_m$ is the number of messenger generations equal to 1 in the minimal model. 
The messenger threshold functions $g(x)$ and $f(x)$ are:
\be
g(x) = \frac{1+x}{x^2}\log(1+x) + (x \to -x)\,, 
\ee
\be
f(x) = g(x) - \frac{2(1+x)}{x^2}[Li_2(\frac{x}{1+x}) - 
\frac{1}{4}Li_2(\frac{x}{1+x})] + (x \to -x).
\ee
It is worthwhile to note, that $g(x) \simeq f(x) \simeq 1$
unless M is close to $\Lambda$ and both functions go to an infinity 
as M equals to $\Lambda$.

The formulae for gaugino and sfermion masses at the 
messenger scale M set up the boundary 
conditions for renormalization group equations (RGE) used to calculate 
the low--energy spectrum. 
In our procedure (see also \cite{guts}) we firstly evolve all the gauge
and Yukawa couplings up to the scale M using the 2--loop standard model 
RGE's below the mass threshold for SUSY particles (initially set up 
to $M_{SUSY} = 1\tev$) and MSSM RGE's above that scale. 
When the couplings reach the scale M we find the values of gaugino and 
scalar masses and perform the RGE evolution of all the quantities
back to $\mz$. 
It is well known, that running of $m_{H_u}^2$ is dominated by negative
contribution from the top Yukawa coupling, which drives this parameter
to a negative value at some scale and  causes  dynamic breaking
of the electroweak symmetry (EWSB). 
This mechanism additionally allows to express some GUT--scale free 
parameters in terms of low--energy ones. 
	
The tree--level Higgs potential has the form:
\be
V_0 = m_1^2\vert H_d^0\vert^2 + m_2^2\vert H_u^0\vert^2 +
m_3^2(H_d^0H_u^0 +
h.c.) + \frac{g_1^2+g_2^2}{8}(\vert H_d^0\vert^2 - \vert
H_u^0\vert^2)^2,
\label{vtree:eq}
\ee
where $m_{1,2}^2\equiv m_{H_{d,u}}^2 + \mu^2, m_3^2\equiv B\mu$,
and phases of fields are chosen so that $m_3^2<0$.
Due to the Q--scale dependence of soft Higgs parameters, $V_0$ 
depends strongly on Q. 
Thus, minimization of $V_0$ can lead to different v.e.v.'s 
$(v_u, v_d)$ for different choices of Q. 
This is the reason why it is much more convenient to minimize 
the full one--loop Higgs effective potential. 
The procedure leads to the set of two equations:
\bea
|\mu|^2+\frac{\mz^2}{2} & = & 
 \frac{(m^2_{H_d}+\Sigma_d)-(m^2_{H_u}+\Sigma_u)\tan^2\beta}
{\tan^2\beta -1}, 
   \label{mincona} \\ 
   \label{minconb}  
  \sin 2\beta & = & 
  \frac{-2B\mu}{(m^2_{H_u}+\Sigma_u)+(m^2_{H_d}+\Sigma_d)+2|\mu|^2},
\eea
where $\Sigma_u, \Sigma_d$ are given in e.g. \cite{finitehiggs}.
In order to minimize the stop contribution to the finite corrections 
we equal the minimization scale $Q_{min}$  to the geometric mean 
of stop masses.

EWSB causes mixing among many particles. 
In particular, mixing in the gaugino sector results in four 
physical neutralinos  $\chi_{i} (i = 1, 2, 3, 4)$,
\begin{equation}
\chi _{i}=\sum_{j=1}^{4}N_{ij}\psi _{j}\,.
\end{equation}
In this equation the matrix $N$ diagonalizing the matrix 
$M_{\chi }$, is real and orthogonal. 
Thus, for real $N_{ij}$ neutralino masses are either positive or
negative. 
If necessary, a negative mass can always be made positive by a 
redefinition of the relevant mixing coefficients $N_{ij} \to iN_{ij}$. 

Similar mixing appears in the slepton and squark sector. 
After calculation of the scalar particles masses and obtaining 
tree level values for $\mu$ and $B\mu$, one is able to set up 
thresholds in RGE's for gauge and Yukawa couplings. 
Repeating the procedure by running everything up to the M scale 
and setting the M--scale conditions for soft parameters again, 
we proceed back to the $\mz$ scale. 
At this point we calculate new mass eigenstates, 
run everything up to $Q_{min}$, minimize the 
one--loop corrected Higgs potential and perform
RGE's run of all the quantities to the $\mz$ scale. 
We iterate the procedure to obtain stable values of $\mu$ and $B\mu$. 

Following the above scheme one can calculate the low--energy values 
of supersymmetric parameters appearing in the 
effective Lagrangian (\ref{susy.2}). 
Some of them can be restricted using experimental data 
for the non-observability of exotic nuclear processes. 
For example limits on $\lambda_{111}'$ can be deduced from 
the non--observability of $0 \nu\beta\beta$.
In this case the half--life for the neutrinoless double beta decay 
regarding all possibilities of hadronization of quarks can be written 
in the form: 
\be
\big[ T_{1/2}^{0\nu}(0^+ \rightarrow 0^+) \big]^{-1}
~=~G_{01} \left | \eta_{T} {\cal M}_{\tilde q}^{2N}
 +  (\eta_{PS} - \eta_{T})~ {\cal M}_{\tilde f}^{2N} + \frac{3}{8}
 (\eta_{T} +  \frac{5}{8} \eta_{PS})~ {\cal M}^{\pi N} \right |^2\,,
\label{susy.3}
\ee
where ${\cal M}^{2N}_{\tilde q}, {\cal M}^{2N}_{\tilde f}$ and 
${\cal M}^{\pi N}$ are defined in Ref. \cite{koval2} and
$G_{01}$ is the standard phase space factor.

Using appropriate values for the nuclear matrix elements 
${\cal M}^{2N}_{\tilde q}, {\cal M}^{2N}_{\tilde f},
{\cal M}^{\pi N}$ and the one--pion and two--pion contribution 
which dominate 2--N mechanism \cite{koval2} we could obtain the 
neutrinoless double beta decay half--life as a function a few 
supersymmetric parameters entering formula (\ref{susy.2}) only. 
Our procedure limits the number of these free parameters
to: $\Lambda, M, \tan\beta, sign(\mu)$ and $N_m$.
As the loop diagrams with messenger fields do not affect 
the A--terms considerably, we can equal the common soft SUSY breaking 
parameter $A_0$ to zero at the M--scale. 

Setting the upper limit on the half--life of the neutrinoless
double beta decay in $^{76}$Ge to $1.1*10^{25} y$ \cite{heidelberg-moscow}
we are able to study constraints imposed by experiment on the R--parity 
breaking parameter  $\lambda_{111}'$ as a function of the above mentioned 
free parameters. 

We have found extremely weak dependence of $\lambda_{111}'$ on
$ M, \tan\beta$ and $sign(\mu)$
and much more pronounced influence of the $\Lambda$ parameter. 
For the  quantities commonly used in literature: $\lambda_{111}'$, 
 $\frac{\lambda_{111}'}{(\msq/100GeV)}$ and $\frac{\lambda_{111}'}
{(\msq/100GeV)^2(\mgluino/100GeV)^{1/2}}$
such a dependence is shown in Fig. 1 (a)-(c). 
The nuclear matrix elements used in these calculations were 
obtained for the bag model. 
It is worth noting that the strongest constraints come from 
the minimal scenario of GMSB ($N_m$ = 1). 

\begin{figure}
\parbox{13cm}{
\epsfxsize=13cm
\epsfbox{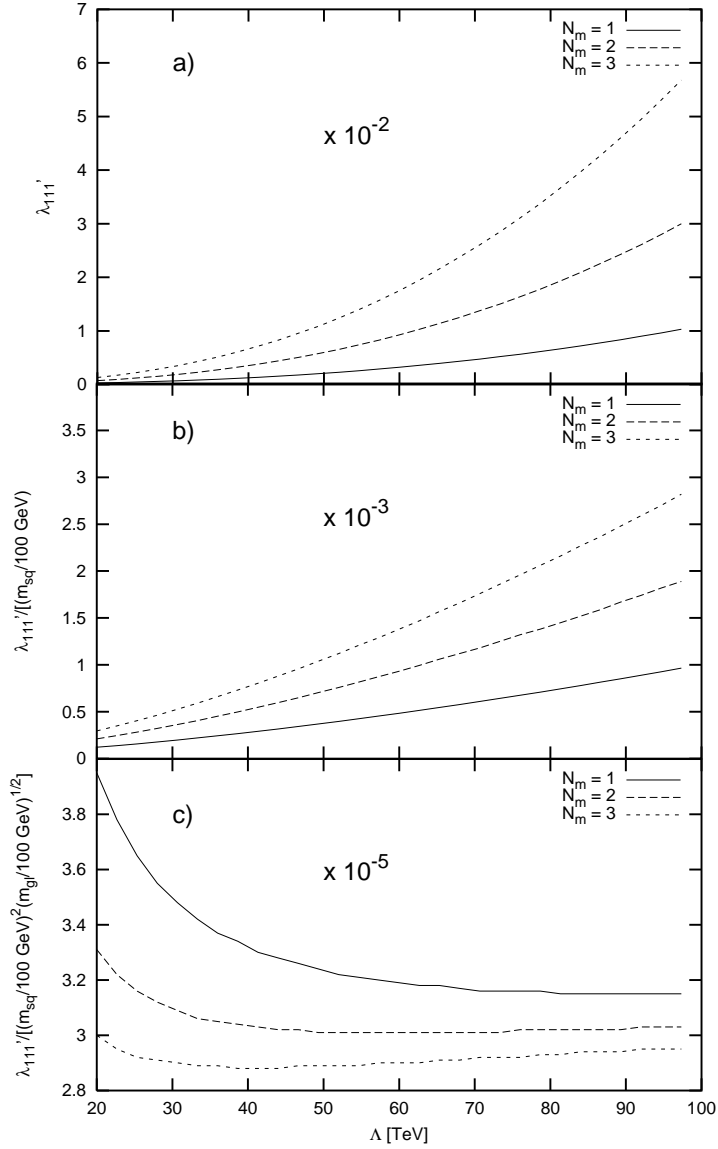}}
\caption{Dependence of different representations of the 
$\lambda_{111}'$ parameter on $\Lambda$. Three curves for 
different numbers of the messenger multiplets ($N_m = 1, 2, 3$)
are shown. Other free parameters are fixed as follows:
$ M = 100 \tev,  \tan\beta = 20, sign(\mu) = +1$. For
details see the text.}
\end{figure}

In conclusions, we have applied the experimental limits on 
the neutrinoless double beta decay of $^{76}$Ge
to analysis of the R--parity non--conservation phenomenology
within the GMSB theory. 
We have shown that this process imposes strong limits on the 
lepton number violating constant $\lambda_{111}'$.
We have also found that $\lambda_{111}'$ depends 
significantly on $\Lambda$ only. 
Moreover, this dependence is stabilizing in the case of 
$\frac{\lambda_{111}'}{(\msq/100GeV)^2(\mgluino/100GeV)^{1/2}}$
for $\Lambda > 70 \tev$ which allows to estimate the lepton number
violating constant:
 \be
 \frac{\lambda_{111}'}{(\msq/100GeV)^2(\mgluino/100GeV)^{1/2}} <
 3.2*10^{-5}\,.
\label{limit}
 \ee
At this point we would like to compare our results with the constraints 
on the R--parity broken MSSM coming from other processes.
The limits obtained from considering the charged current universality 
violation impose the folowing bound:$\lambda'_{111} \leq 0.03 
(\frac{m_{{\tilde d}_R}}{100 GeV})$ \cite{barger}.
The above limit has been found within the gravity mediated MSSM. 
One can observe that in the GMSB case constraints are
almost 2 orders of magnitude stronger. 

Comparison with other estimates \cite{koval1,wodecki} shows
that limit (\ref{limit}) is approximately one order of magnitude
stronger. These results may be helpful for planning
future searches for R-parity breaking signatures in 
different experiments.

The presented approach serves as a consistent and free of
{\it ad hoc} assumptions about mutual dependence of model
parameters one. 
To the best our knowledge we are first applying the gauge 
mediated supersymmetry breaking mechanism to constrain the 
R--parity breaking through the neutrinoless double beta decay.

We are grateful to F. Simkovic for providing us with the nuclear 
matrix elements prior to their publication. 
This work was supported in part by the State Committee for Scientific 
Researches (Poland).   



\end{document}